\documentclass[12pt]{article}
\usepackage{graphicx,epsf,amsmath}

\def\Dslash{\rlap{\hspace{0.10cm}/}{D}}

\newcommand\arnps[3]{Ann.\ Rev.\ Nucl.\ Part.\ Sci.\ {\bf #1} (#2) #3} 

\newcommand\ijmpa[3]{Int.\ J.\ Mod.\ Phys.\ A {\bf #1} (#2) #3}
\newcommand\jhep[3]{J.\ High Ener.\ Phys.\ {\bf #1} (#2) #3}
\newcommand\npb[3]{Nucl.\ Phys.\ B {\bf #1} (#2) #3}

\newcommand\plb[3]{Phys.\ Lett.\ B {\bf #1} (#2) #3}
\newcommand\prd[3]{Phys.\ Rev.\ D {\bf #1} (#2) #3}
\newcommand\prep[3]{Phys.\ Rep.\ {\bf #1} (#2) #3}
\newcommand\prl[3]{Phys.\ Rev.\ Lett.\ {\bf #1} (#2) #3}

\newcommand{\hepph}[1]{{\tt hep-ph/#1}}

\newcommand{\ibid}[3]{ibid.\ {\bf #1} (#2) #3} 

\newcommand{\mbp}{m_b^{\text{pole}}}
\newcommand{\mbPS}{m_b^{\text{PS}}}
\newcommand{\mbUp}{m_b^{\text{1S}}}

\begin{document}

\begin{titlepage}

\begin{flushright}
CLNS~01/1737\\
{\tt hep-ph/0105217}\\[0.2cm]
May 21, 2001
\end{flushright}

\vspace{1.5cm}
\begin{center}
\Large\bf\boldmath
Improved Determination of $|V_{ub}|$ from Inclusive Semileptonic 
$B$-Meson Decays
\unboldmath
\end{center}

\vspace{0.5cm}
\begin{center}
Thomas Becher and Matthias Neubert\\[0.1cm]
{\sl Newman Laboratory of Nuclear Studies, Cornell University\\
Ithaca, NY 14853, USA}
\end{center}

\vspace{1.5cm}
\begin{abstract}
\vspace{0.2cm}\noindent 
We reduce the perturbative uncertainty in the
determination of $|V_{ub}|$ from inclusive semileptonic $B$ decays by
calculating the rate of $B\to X_u\,l\,\nu$ events with dilepton
invariant mass $\sqrt{q^2}>M_B-M_D$ at subleading order in the hybrid
expansion, and to next-to-leading order in renormalization-group
improved perturbation theory. We also resum logarithmic corrections to
the leading power-suppressed contributions. Studying the effect of
different $b$-quark mass definitions we find that the branching ratio
after the cut is $\mbox{Br}(B\to X_u\,l\,\nu)=(20.9\pm 4.0)\,|V_{ub}|^2$, 
where the dominant error is due to the uncertainty 
in the $b$-quark mass. This implies that $|V_{ub}|$ can be determined
with a precision of about 10\%.
\end{abstract}

\end{titlepage}

\section{Introduction}

The total inclusive, semileptonic $B\to X_u\,l\,\nu$ decay rate admits a 
clean theoretical analysis based on the heavy-quark expansion. 
The optical theorem is used to relate the decay rate to the forward 
matrix element of the product of two hadronic currents, which is 
evaluated using the operator product expansion \cite{Chay,Bigi,MaWe}. 
The total rate is presently known 
including corrections of order $\alpha_s^2$ \cite{Ritb} and 
$\Lambda^2/m_b^2$ (with $\Lambda$ a typical hadronic scale). An accurate 
knowledge of the total rate does not immediately translate into a 
precise determination of $|V_{ub}|$, since experimentally one has to 
impose cuts on the kinematical variables in order to discriminate the 
$b\to u$ transitions against the much larger $B\to X_c\,l\,\nu$ 
background. The first inclusive determination of $|V_{ub}|$ was based on 
a measurement of the charged-lepton energy spectrum in the endpoint 
region, which is kinematically forbidden for decays with a charm meson 
in the final state. This cut eliminates about 90\% of the signal. It is
difficult to reliably calculate the remaining event 
fraction, since the operator product expansion breaks down in this part 
of the phase space and must be replaced by a twist expansion 
\cite{shape,Russ}.

Recently, it was shown that the situation is improved when a cut 
$q^2>q_0^2$ with $q_0^2\ge(M_B-M_D)^2$ is applied to the dilepton 
invariant mass \cite{Bauer:2000xf}. This keeps about 20\% of the signal
events, and the fraction of the remaining events can be calculated in a 
model-independent way using the operator product expansion. After the 
cut, the typical momentum flow $\mu_c$ through the hadronic tensor is of 
order the charm-quark mass or less, so that the problem at hand 
involves the three scales $m_b>\mu_c>\Lambda$. In the standard 
heavy-quark expansion, the charm and the $b$-quark masses are treated on 
the same footing, but it is known that the convergence of this expansion 
at the charm scale can be poor. In this letter we follow the strategy 
proposed in \cite{Neubert:2000ch} and perform a two-step expansion (the 
``hybrid expansion'') in the ratios $\mu_c/m_b$ and $\Lambda/\mu_c$. The 
first step is to integrate out the degrees of freedom above the scale 
$\mu_c$ and to sum up logarithms of the form $\alpha_s\ln(\mu_c/m_b)$ 
using the renormalization group (RG). This is achieved by expanding the 
weak currents and the QCD Lagrangian to order $1/m_b$ in the heavy-quark 
effective theory (HQET) \cite{review}, and calculating the Wilson 
coefficients of the pertinent operators at next-to-leading order (NLO) 
in RG-improved perturbation theory. Whereas the Wilson coefficients of 
the leading-order currents \cite{Ji:1991pr,Broadhurst:1991fz} and of the 
$O(1/m_b)$ corrections to the HQET Lagrangian 
\cite{Amoros:1997rx,Czarnecki:1997dz} were known for a long time, the 
NLO calculation of the coefficients entering the expansion of the 
currents at order $1/m_b$ was completed only recently 
\cite{FaGr,Neubert:1994za,Amoros:1998xr,Becher:2000nm}.

In the present work, we perform the second step in the hybrid expansion 
and construct the operator product expansion in the resulting low-energy
effective theory. We calculate the decay rate 
$\Gamma(B\to X_u\,l\,\nu)|_{q^2>q_0^2}$ at subleading order in the ratio
$\mu_c/m_b$ and to NLO in RG-improved perturbation theory. We also resum 
leading logarithmic corrections to the power corrections of order 
$(\Lambda/\mu_c)^2$. In contrast with previous analyses 
\cite{Bauer:2000xf,Neubert:2000ch} we calculate the decay rate directly, 
rather than calculating the fraction of events that passes the $q^2$-cut 
and then multiplying this fraction with an independent result for the 
total semileptonic rate. Since the characteristic scale of the 
total rate is significantly higher than (and parametrically different 
from) the typical hadronic scale $\mu_c$ after the $q^2$-cut, the
theoretical error cannot be reduced by separating the two calculations.
Based on a detailed analysis of the residual renormalization-scale 
dependence and the sensitivity to the definition of the
$b$-quark mass, we conclude that the $B\to X_u\,l\,\nu$ branching ratio
after the dilepton invariant-mass cut can be calculated reliably, 
allowing for a determination of $|V_{ub}|$ at the 10\% level.

\section{Construction of the hybrid expansion}

A cut $q^2>q_0^2$ on the dilepton invariant mass squared 
$q^2=(p_l+p_\nu)^2$ implies
\begin{equation}
   M_X\le E_X\le \frac{1}{2M_B} \left( M_B^2+M_X^2-q_0^2 \right) \,,
    \qquad
   M_\pi^2\le M_X^2\le \Big( M_B-\sqrt{q_0^2} \Big)^2
\end{equation}
for the hadronic invariant mass $M_X$ and energy $E_X$. For 
$q_0^2=(M_B-M_D)^2$ both variables vary between 
$M_\pi$ and $M_D$, and thus no charmed hadrons are allowed in the final
state. If the cutoff is chosen higher (as may be required for 
experimental reasons), the maximal values of $M_X$ and $E_X$ become less 
than the $D$-meson mass. The kinematical variables that enter the 
partonic calculation of the decay rate are the parton invariant mass and 
energy, $\sqrt{p^2}$ and $v\cdot p$, where $p=m_b\,v-q$ with $v$ being 
the velocity of the $B$-meson. 
It is convenient to work with the dimensionless quantities 
$\xi=p^2/(v\cdot p)^2$ and $z=2v\cdot p/m_b$, the phase space for which 
is
\begin{equation}\label{eq:zmax}
   0\le\xi\le 1 \,, \qquad
   0\le z\le z_{\rm max} 
    = \frac{2}{\xi} \Big( 1-\sqrt{1-2\delta\,\xi} \Big) \,. 
\end{equation}
While $\xi$ is of order unity, independent of the value of $q_0^2$, the 
quantity $z$ is parametrically suppressed by 
\begin{equation}
   \delta = \frac{\mu_c}{m_b} \,, 
   \qquad\text{with}\qquad 
   \mu_c = \frac{m_b^2-q_0^2}{2 m_b} \,.
\end{equation}
With the optimal value $q_0^2=(M_B-M_D)^2\simeq (m_b-m_c)^2$ the
characteristic scale of the parton momenta is $\mu_c\sim m_c$.

The strong-interaction dynamics of the inclusive decay $B\to X_u\,l\,\nu$
is encoded in the imaginary part of the hadronic tensor $T_{\mu\nu}$
defined as 
\begin{eqnarray}\label{Tprod}
   T_{\mu\nu}(p,v) &=& \frac{1}{i} \int d^4 x\,e^{i(p-m_b v)\cdot x}\,
    \frac{1}{2M_B}\,
    \langle B(v)|\,\text{\bf T} \left\{ J_\mu^\dag(x), J_\nu(0) 
    \right\}|B(v)\rangle \nonumber\\
   &=& \left( p_\mu v_\nu + v_\mu p_\nu - g_{\mu\nu}\,v\cdot p 
    - i\epsilon_{\mu\nu\alpha\beta}\,p^\alpha v^\beta \right)\,T_1
    \nonumber\\
   &&\mbox{}- g_{\mu\nu}\,T_2 + v_\mu v_\nu\,T_3
    + \left( p_\mu v_\nu + v_\mu p_\nu \right)\,T_4
    + p_\mu p_\nu\,T_5 \,,
\end{eqnarray}
where the scalar functions $T_i$ depend on $p^2$ and $v\cdot p$, or 
equivalently, on $\xi$ and $z$. The weak current 
$J_\mu=\bar u\gamma_\mu(1-\gamma_5)b$ mediates the $b\to u$ transition. 
To separate the long- and short-distance physics one usually performs 
the operator product expansion of the hadronic tensor and calculates the 
corresponding Wilson coefficient functions in perturbation theory. 
However, since in the present case the characteristic scale $\mu_c$ of 
the momentum flow through the operator product is lower than the 
$b$-quark mass, the perturbative expansion of the Wilson coefficients
involves logarithms of the ratio $\mu_c/m_b$. To improve the quality of 
the perturbative results these logarithms should be summed. This can be 
achieved by absorbing the physics above the scale $\mu_c$ into the 
couplings of a low-energy effective theory. 

The relevant effective theory is the HQET, and the expansion of the weak
current at NLO reads \cite{FaGr,Neubert:1994za}
\begin{equation}\label{jexp}
   J_\mu = \sum_i C_i(\mu)\,O_i + \frac{1}{2 m_b} \sum_j B_j(\mu)\,Q_j
   + O(1/m_b^2) \,.
\end{equation}
The two operators contributing at leading order are 
\begin{equation}
   O_1 = \bar u_L\gamma_\mu h_v \,, \qquad
   O_2 = \bar u_L v_\mu h_v \,,
\end{equation}
where $\bar u_L\equiv\bar u(1+\gamma_5)$, and $h_v$ is the effective
heavy-quark spinor of the HQET. At NLO in $1/m_b$, the operator basis 
contains ten independent operators. Six of these are local operators 
involving a gauge-covariant derivative on one of the quark fields:
\begin{align}
   Q_1 &= \bar q_L \gamma_\mu i\Dslash h_v \,, &
   Q_2 &= \bar q_L v_\mu i\Dslash h_v \,, & 
   Q_3 &= \bar q_L iD_\mu h_v \,, \nonumber\\
   Q_4 &= \bar q_L (iv\!\cdot\!D)^\dagger\gamma_\mu h_v \,, & 
   Q_5 &= \bar q_L (iv\!\cdot\!D)^\dagger v_\mu h_v \,, &
   Q_6 &= \bar q_L (i D_\mu)^\dagger h_v \,.
\end{align}
The remaining four involve time-ordered products of the leading-order 
currents with the operators $O_{\rm kin}=\bar h_v (iD)^2 h_v$ and 
$O_{\rm mag}=\frac{g_s}{2}\,\bar h_v\sigma_{\mu\nu} G^{\mu\nu} h_v$ 
appearing at order $1/m_b$ in the HQET Lagrangian:
\begin{equation}
   Q_{7,8} = i\int d^4x\,\mbox{\bf T}\,\{O_{1,2}(0),O_{\rm kin}(x)\} \,,
    \qquad
   Q_{9,10} = i\int d^4x\,\mbox{\bf T}\,\{O_{1,2}(0),O_{\rm mag}(x)\} \,. 
\end{equation}
These nonlocal operators are present because we work with eigenstates of 
the lowest-order HQET Lagrangian, so that all dependence on the 
heavy-quark mass is explicit. It is convenient to treat the operators 
$Q_{7,\dots,10}$ on the same footing as the local operators 
$Q_{1,\dots,6}$, because the nonlocal operators mix into the local ones 
under renormalization. We have recently completed the calculation of the 
two-loop anomalous dimension matrix for the operators $Q_j$ and solved 
the RG equations for their coefficients $B_j(\mu)$ \cite{Becher:2000nm}. 

When the expansion (\ref{jexp}) is inserted into the operator product
in (\ref{Tprod}), all dependence on the $b$-quark mass becomes explicit 
and is contained in the Wilson coefficients $C_i$, $B_j$ and the 
$1/m_b$ prefactor of the higher-dimension operators. In 
the next step, we perform the operator product expansion and obtain 
forward $B$-meson matrix elements of local operators of the form 
$\langle B(v)|\,\bar h_v\dots h_v\,|B(v)\rangle$ in the low-energy 
theory. To order $(\Lambda/\mu_c)^2$, these matrix elements can be 
expressed in terms of the parameters \cite{FaNe}
\begin{equation}
   \lambda_1 = \frac{1}{2 M_B}\,\langle B|\,O_{\rm kin}\,|B\rangle \,,
    \qquad
   \lambda_2(\mu) 
   = \frac{1}{6 M_B}\,\langle B|\,O_{\rm mag}\,|B\rangle \,.
\end{equation}
The anomalous dimension of the kinetic operator vanishes because of
reparameterization invariance \cite{LuMa}. Up to higher orders in the 
heavy-quark expansion, $\lambda_2$ is determined from the mass splitting
$\lambda_2(m_b)\simeq\frac{1}{4}(M_{B^*}^2-M_B^2)\simeq 0.12$\,GeV$^2$. 

By dimensional analysis, the contributions from time-ordered products 
of higher-dimension operators in (\ref{jexp}) are suppressed by powers 
of $\delta=\mu_c/m_b$. At NLO in $\delta$, it suffices to evaluate 
time-ordered products with operator insertions of the type $\{O_i,O_j\}$ 
or $\{O_i,Q_j\}$. 
The decay rate is then obtained by contracting the result for the 
hadronic tensor with the lepton tensor and integrating over the portion 
of phase space with $q^2>q_0^2$:
\begin{equation}\label{eq:phaseint}
   \Gamma(\delta)\equiv \Gamma(B\to X_u\,l\,\nu)\big|_{q^2>q_0^2}
   = \int\limits_0^1\!d\xi \int\limits_0^{z_{\rm max}}\!dz\,
    \frac{d^2\Gamma}{d\xi\,dz}
   = \frac{1}{2\pi i} \oint\limits_{|\xi|=1}\!d\xi
    \int\limits_0^{z_{\rm max}}\!dz\,T(\xi,z) \,,
\end{equation}
where $z_{\rm max}=2\delta+\delta^2\xi+O(\delta^2)$ from (\ref{eq:zmax}).
Using the fact that the hadronic tensor is an analytic function of $\xi$ 
apart from discontinuities on the positive real axis, we have written the
expression for the rate $\Gamma(\delta)$ as a contour integral in the 
complex $\xi$-plane. The integrand of the contour integral is 
\cite{Fulvia}
\begin{eqnarray}\label{eq:diffrate}
   T(\xi,z) &=& \frac{\Gamma_0 m_b z^3}{16} \sqrt{1-\xi}\,
    \bigg\{ m_b z \Big[ 12-8z-(4-3z) z\,\xi \Big]\,T_1 
    + 6(4-4z+z^2\xi)\,T_2 \nonumber\\
   &&\hspace{3.1cm}\mbox{}+ 2 z^2(1-\xi)(T_3 + 2 m_b\,T_4 + m_b^2\,T_5)
    \bigg\} \,,
\end{eqnarray}
where 
\begin{equation}
   \Gamma_0 = \frac{G_F^2 |V_{ub}|^2 m_b^5}{192\pi^3}
\end{equation}
is the tree-level decay rate.

\begin{figure}[t]
\begin{center}
\includegraphics[width=0.9\textwidth]{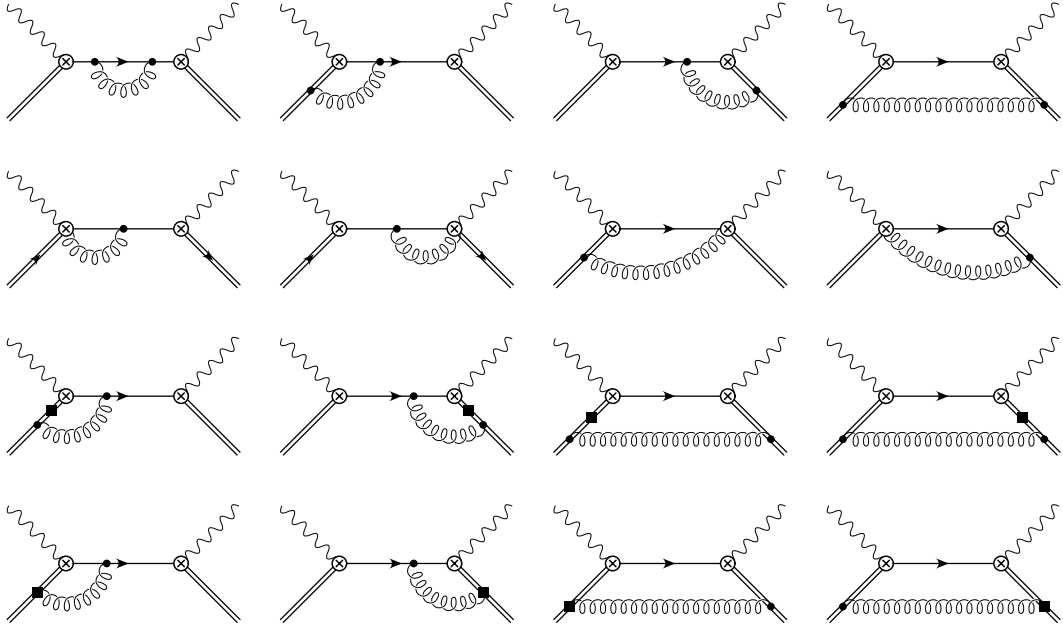}
\vspace{0.2cm}\\
\parbox{14cm}{\caption{\label{fig:graphs}
One-loop graphs contributing to the hadronic tensor. The crossed circles
represent insertions of the operators $O_i$ or $Q_j$. The solid 
squares in the last eight diagrams denote insertions of $O_{\rm kin}$ 
or $O_{\rm mag}$.}}
\end{center}
\end{figure}

At NLO in RG-improved perturbation theory, the matrix elements of the 
hadronic tensor must be evaluated including corrections of order 
$\alpha_s$ in the low-energy theory. These are obtained by evaluating 
the contributions of all physical cuts of the diagrams shown in 
Figure~\ref{fig:graphs}. These diagrams contain infrared divergences for 
$\xi\to 0$ from soft and collinear gluons, which cancel once the 
differential rate is integrated over phase space, because the rate 
$\Gamma(\delta)$ is an infrared-safe quantity. The 
contour-integral representation in (\ref{eq:phaseint}) avoids the 
singularity at $\xi=0$. After integrating over the loop momentum, the 
function 
$T(\xi,z)$ is given by an integral over Feynman parameters. At this point
we exchange the order of integration and integrate over the phase space 
first. The $z$-integration is trivial, because the dependence of the 
loop integrals on $z$ is fixed by their mass dimension. Note 
that the upper limit $z_{\rm max}$ of the integration depends on $\xi$. 
After expanding to NLO in $\delta$, the 
$\xi$-integrals can all be reduced to
\begin{equation}
   \big\{ G_\alpha(a)\,;H_\alpha(a) \big\}
   = \frac{1}{2\pi i}\oint\limits_{|\xi|=1}\!d\xi\,
   \frac{\sqrt{1-\xi}}{(a-\xi)^\alpha}\,
   \big\{ 1\,;\frac{1}{\xi}\, \big\} \,,
\end{equation}
where the denominator $(a-\xi)^\alpha$ with 
$a\ge 0$ is the result of the Feynman parameterization. Some of the
diagrams have a pole at $\xi=0$ due to the propagation of a single
$u$-quark. The integral $G_\alpha(a)$ evaluates to
\begin{equation}
   G_\alpha(a) 
   = - \frac{\sqrt\pi}{2\Gamma(\alpha)\Gamma(\frac52-\alpha)}\,
   \frac{\theta(1-a)}{(1-a)^{\alpha-\frac32}} \,.
\end{equation}
The second integral picks up contributions for $a<1$ and $a>1$. It is
needed only for $\alpha=1+\epsilon$ and 
$\alpha=\epsilon$ (with $\epsilon=2-d/2$ the dimensional regulator). We 
find
\begin{eqnarray}
   H_{1+\epsilon}(a) &=& \frac{\theta(1-a)}{a}\,\bigg\{
    1 - \sqrt{1-a}\,\Big[ 1 - \epsilon\ln4(1-a) \Big] 
    - 2\epsilon \ln(1+\sqrt{1-a}) + O(\epsilon^2) \bigg\} \nonumber\\
   &&\mbox{}+ \frac{\theta(a-1)}{a^{1+\epsilon}} \,.
\end{eqnarray} 
We have not expanded the result for $a>1$, since we will have to 
integrate it from $a=1$ to $\infty$ when performing the integrals over 
Feynman parameters. The corresponding expression for the function 
$H_\epsilon(a)$ can be obtained by integrating the above result with 
respect to $a$.

We now present the results of our calculation. We expand the rate 
$\Gamma(\delta)$ in powers of $\delta$ and define
\begin{equation}\label{hybrid}
   \Gamma(\delta) = 16\Gamma_0\,\delta^3 \left( \widehat\Gamma_3
   + \delta\,\widehat\Gamma_4 + \delta^2\,\widehat\Gamma_{\ge 5} \right)
   ,
\end{equation}
where the dimensionless expansion coefficients $\widehat\Gamma_n$ 
collect terms of different order in the hybrid expansion. The 
leading-order contribution $\widehat\Gamma_3$ was calculated in 
\cite{Neubert:2000ch}; the NLO term $\widehat\Gamma_4$ is the main 
result of the present work. We find that
\begin{eqnarray}\label{G3G4}
   \widehat\Gamma_3 &=& C_1(\mu)^2\,F_3[C_1,C_1] \,, \nonumber\\
   \widehat\Gamma_4 &=& \sum_{i=1,2}\,C_1(\mu)\,C_i(\mu)\,F_4[C_1,C_i]
    + \sum_{j=4,6,7,9}\,C_1(\mu)\,B_j(\mu)\,F_4[C_1,B_j] \,,
\end{eqnarray}
where ($C_F=4/3$ is a color factor)
\begin{eqnarray}\label{Fres}
   F_3[C_1,C_1] &=& 1 + \frac{C_F\alpha_s(\mu)}{4\pi} \left(
    6\ln\frac{\mu}{2\mu_c} + 19 - \frac{8\pi^2}{3} \right)
    - \frac32\,\frac{\lambda_2(\mu)}{\mu_c^2} + \dots \,, \nonumber\\
   F_4[C_1,C_1] &=& - \bigg\{ 1 + \frac{C_F\alpha_s(\mu)}{4\pi} \left(
    6\ln\frac{\mu}{2\mu_c} + \frac{29}{2} - \frac{8\pi^2}{3} \right)
    \bigg\} + \frac32\,\frac{\lambda_1+3\lambda_2(\mu)}{\mu_c^2}
    + \dots \,, \nonumber\\
   F_4[C_1,C_2] &=& \frac12 + \dots \,, \nonumber\\
   F_4[C_1,B_4] &=& \frac34\,\bigg\{ 1 + \frac{C_F\alpha_s(\mu)}{4\pi} 
    \left( 6\ln\frac{\mu}{2\mu_c} + \frac{37}{2} - \frac{8\pi^2}{3} 
    \right) \bigg\} - \frac34\,\frac{\lambda_2(\mu)}{\mu_c^2}
    + \dots \,, \nonumber\\
   F_4[C_1,B_6] &=& \frac14\,\bigg\{ 1 + \frac{C_F\alpha_s(\mu)}{4\pi}
    \left( 6\ln\frac{\mu}{2\mu_c} + \frac{39}{2} - \frac{8\pi^2}{3} 
    \right) \bigg\} + \frac34\,\frac{\lambda_2(\mu)}{\mu_c^2}
    + \dots \,, \nonumber\\
   F_4[C_1,B_7] &=& \frac{C_F\alpha_s(\mu)}{4\pi}
    \left( 6\ln\frac{\mu}{2\mu_c} + \frac92 \right)
    - \frac32\,\frac{\lambda_1}{\mu_c^2} + \dots \,, \nonumber\\
   F_4[C_1,B_9] &=& \frac13\,\frac{C_F\alpha_s(\mu)}{4\pi}
    \left( 6\ln\frac{\mu}{2\mu_c} + \frac{25}{2} \right)
    - \frac92\,\frac{\lambda_2(\mu)}{\mu_c^2} + \dots \,.
\end{eqnarray}
Note that the Wilson coefficient $C_2$ starts at NLO in $\alpha_s$, so 
there is no need to include $O(\alpha_s)$ corrections to the quantity 
$F_4[C_1,C_2]$. The coefficients $B_7=C_1\,C_{\rm kin}$ and 
$B_9=C_1\,C_{\rm mag}$ are determined in terms of products of 
coefficients appearing in the leading-order expansion of the currents 
and in the HQET Lagrangian. $C_{\rm mag}$ is the coefficient of the 
chromo-magnetic operator $O_{\rm mag}$, and $C_{\rm kin}=1$ is the 
coefficient of the kinetic operator $O_{\rm kin}$. The remaining 
coefficients, $B_4$ and $B_6$, were calculated at NLO in 
\cite{Becher:2000nm}. The Wilson coefficients, as well as the matrix 
elements in the low-energy theory, depend on the choice of the 
renormalization scheme. The scheme dependence cancels in the 
result for the decay rate $\Gamma(\delta)$. In \cite{Becher:2000nm} we 
have considered a class of renormalization schemes parameterized by the 
quantity $\Delta_{\rm RS}=2\ln(\mu_{\overline{\rm MS}}/\mu_{\rm RS})$. 
Our results (\ref{Fres}) refer to the $\overline{\rm MS}$ renormalization
scheme (with anticommuting $\gamma_5$). The corresponding results in a 
different scheme RS are obtained by replacing
\begin{equation}
   \ln\frac{\mu_{\overline{\rm MS}}}{2\mu_c}
   \to \ln\frac{\mu_{\rm RS}}{2\mu_c} + \frac{\Delta_{\rm RS}}{2} \,.
\end{equation} 
That way, it can be checked that the scheme-dependent terms proportional
to $\Delta_{\rm RS}$ indeed cancel in the final result.

The above expressions are given in terms of the $b$-quark pole mass 
$m_b\equiv\mbp$, which is the expansion parameter that enters in the 
HQET. However, this mass parameter is plagued by infrared renormalon 
ambiguities, and it is better to eliminate it from the final result for 
the decay rate. It is well known that the convergence of the perturbative 
series of inclusive decay rates can be improved by replacing the pole 
mass in favor of a low-scale subtracted quark mass, which is obtained 
from the pole mass by removing a long-distance contribution
proportional to a subtraction scale $\mu_f\sim\mbox{several}\times
\Lambda$ \cite{PSmass,1Smass,kinmass,Martin}. In 
addition to $m_b$, the characteristic scale $\mu_c=(m_b^2-q_0^2)/(2m_b)$ 
of the hybrid expansion and the associated parameter $\delta=\mu_c/m_b$ 
depend on the definition of the quark mass. It has been shown in 
\cite{Neubert:2000ch} that one should choose $\mu_f=\zeta\mu_c$ with a 
parameter $\zeta=O(1)$ so as not to upset the power-counting in the 
hybrid expansion. Specifically, we use the potential-subtracted (PS) 
mass $\mbPS$ introduced by Beneke \cite{PSmass}. At NLO, the relation 
between the pole 
mass and the PS mass reads\footnote{The NNLO contribution to this 
relation is known, but we will not use it in the present work.}
\begin{equation}\label{eq:PSmass}
   \mbp = \mbPS(\mu_f) + \mu_f\,\frac{4\alpha_s(\mu)}{3\pi}
   \left\{ 1 + \frac{\alpha_s(\mu)}{2\pi} \left[ 
   \beta_0 \left( \ln\frac{\mu}{\mu_f} + \frac{11}{6} \right)
   - 4 \right] + \dots \right\} ,
\end{equation}
where $\beta_0=11-\frac23\,n_f$ is the first coefficient of the $\beta$ 
function. With $\mu_f=\zeta\mu_c$ it follows that
\begin{equation}\label{PSfirst}
   \mbp = \mbPS(\zeta\mu_c) \left[ 1 + \delta\,\zeta\,
   \frac{4\alpha_s(\mu)}{3\pi} + O(\alpha_s^2) \right] .
\end{equation}
From now on, we use the PS mass $m_b\equiv\mbPS(\zeta\mu_c)$ in all our 
equations unless indicated otherwise. The parameters $\mu_c$ and 
$\delta$ are defined with respect to this choice. This results in extra 
terms in the perturbative expansion proportional to $\zeta$. The results 
in the pole scheme can be recovered by setting $\zeta=0$.

In \cite{Bauer:2000xf}, the authors have eliminated the pole mass in 
favor of the so-called Upsilon mass $\mbUp$ \cite{1Smass}, which up to a 
small nonperturbative contribution is one half of the mass of the 
$\Upsilon(1S)$ resonance. The relation between the Upsilon mass and the 
pole mass is
\begin{equation}\label{1Smass}
   \mbp = \mbUp + \frac{2\mbUp\alpha_s(\mu)^2}{9} 
   \left\{ 1 + \frac{\alpha_s(\mu)}{\pi} \left[ 
   \beta_0 \left( \ln\frac{\mu}{\frac43\,\mbUp\alpha_s(\mu)}
   + \frac{11}{6} \right) - 4 \right] + \dots \right\} .
\end{equation}
In higher orders in this expansion the logarithms 
$\ell=\ln[\mu/\frac43\,\mbUp\alpha_s(\mu)]$ appear in the form 
$[\ell^n/n!+\ell^{n-1}/(n-1)!+\dots+1]\sim\exp(\ell)$, which yields a 
factor $[\mu/\frac43\,\mbUp\alpha_s(\mu)]$. Hence, schematically 
the relation between the masses becomes
\begin{equation}
   \mbp\sim \mbUp + \mu\,\frac{\alpha_s(\mu)}{6} 
   \left[ 1 + \mbox{nonlogarithmic terms} \right] \,,
\end{equation}
which shows that despite of the unusual form of (\ref{1Smass}) the 
Upsilon mass is a low-scale subtracted mass in the sense described 
earlier. (One may, however, question whether this argument can 
consistently be applied in low-order calculations.) Based on this 
observation, the authors of \cite{1Smass} have suggested to treat the 
$O(\alpha_s^2)$ term on the right-hand side of (\ref{1Smass}) as an 
$O(\alpha_s)$ correction when the Upsilon mass is used in the context of 
perturbative expansions in the $B$ system. Therefore, in our case we 
should replace $\mbp=\mbUp[1+\frac29\,\alpha_s(\mu)^2+\dots]$ and treat 
the $O(\alpha_s^2)$ term as a term of order $\delta\,\alpha_s(\mu)$, as 
in (\ref{PSfirst}). This means that our formulae apply without 
modification to the Upsilon scheme, provided we set 
$\zeta=\pi\alpha_s(\mu)/(6\delta)$ and consider this to be a parameter 
of order unity.  

Inserting the explicit expressions for the Wilson coefficients and
matrix elements into (\ref{G3G4}), and evaluating the results for 
$N_c=3$ colors and $n_f=4$ light quark flavors, we obtain
\begin{eqnarray}\label{master}
   x^{-12/25}\,\widehat\Gamma_3
   &=& 1 + k_1\,\frac{\alpha_s(m_b)}{\pi}
    + \left( 2\ln\frac{\mu}{2\mu_c} + 4\zeta + k_2 \right)
    \frac{\alpha_s(\mu)}{\pi}
    - \frac32\,x^{9/25}\,\frac{\lambda_2(m_b)}{\mu_c^2} + \dots \,,
    \nonumber\\
   x^{-12/25}\,\widehat\Gamma_4
   &=& - \frac59 + k_3\,\frac{\alpha_s(m_b)}{\pi}
    + \left( \frac89\,\ln\frac{\mu}{2\mu_c} - \frac{116}{27}\,\zeta 
    + k_4 \right) \frac{\alpha_s(\mu)}{\pi} \nonumber\\
   &&\mbox{}+ x^{-9/25}\,\left[ - \frac49
    + k_5\,\frac{\alpha_s(m_b)}{\pi}
    + \left( - \frac29\,\ln\frac{\mu}{2\mu_c} 
    - \frac{64}{27}\,\zeta + k_6 \right) 
    \frac{\alpha_s(\mu)}{\pi} \right] \nonumber\\
   &&\mbox{}+ \ln x\,\left[ \frac{12}{25}
    + k_7\,\frac{\alpha_s(m_b)}{\pi} 
    + \left( \frac{24}{25}\,\ln\frac{\mu}{2\mu_c} 
    + \frac{64}{25}\,\zeta + k_8 \right) 
    \frac{\alpha_s(\mu)}{\pi} \right] \nonumber\\
   &&\mbox{}+ \left( \frac{37}{18} - \frac{12}{25}\,\ln x 
    + \frac{10}{3}\,x^{-6/25} - \frac{97}{18}\,x^{-9/25} \right)
    x^{9/25}\,\frac{\lambda_2(m_b)}{\mu_c^2} + \dots \,,
\end{eqnarray}
where $x=\alpha_s(\mu)/\alpha_s(m_b)$, and we have used that 
$\lambda_2(\mu)=x^{9/25}\,\lambda_2(m_b)$. We have extracted an 
overall factor $x^{12/25}$ from the expressions for the quantities 
$\widehat\Gamma_n$, which reflects the leading-order anomalous 
dimensions of the heavy--light currents in the time-ordered product in 
(\ref{Tprod}). The coefficients $k_i$ summarizing the NLO corrections 
are
\begin{align}
   k_1 &= - \frac{9403}{3750} - \frac{14\pi^2}{225} 
    \simeq -3.122 \,, &
   k_2 &= \frac{23153}{3750} - \frac{62\pi^2}{75}
    \simeq -1.985 \,, \nonumber\\
   k_3 &= \frac{23161}{180000} + \frac{1123\pi^2}{8100}
    \simeq 1.497 \,, &
   k_4 &= \frac{22106}{16875} + \frac{196\pi^2}{675}
    \simeq 4.176 \,, \nonumber\\
   k_5 &= \frac{988}{5625} + \frac{56\pi^2}{2025}
    \simeq 0.449 \,, &
   k_6 &= - \frac{661723}{540000} + \frac{389\pi^2}{900}
    \simeq 3.040 \,, \nonumber\\
   k_7 &= - \frac{18806}{15625} - \frac{56\pi^2}{1875} 
    \simeq -1.498 \,, &
   k_8 &= \frac{45056}{15625} - \frac{248\pi^2}{625} 
    \simeq -1.033 \,.
\end{align}

Since we have solved the RG equation for the Wilson coefficients to 
obtain their running from the $b$-quark mass down to the scale 
$\mu\sim\mu_c$, all logarithms of the ratio $\mu/m_b$ are resummed into 
the running couplings $\alpha_s(\mu)$ and $\alpha_s(m_b)$. Note that 
the expressions (\ref{master}) are scale- and scheme-independent at NLO. 
As a check of our result, we expand $\alpha_s(\mu)$ in terms of 
$\alpha_s(m_b)$ in order to recover the known one-loop expression for 
the decay rate. This yields
\begin{eqnarray}
   \widehat\Gamma_3
   &=& 1 + \frac{\alpha_s}{\pi} \left( 2\ln\frac{m_b}{2\mu_c}
    + 4\zeta + \frac{11}{3} - \frac{8\pi^2}{9} \right) 
    - \frac32\,\frac{\lambda_2}{\mu_c^2} + \dots \,, \nonumber\\
   \widehat\Gamma_4
   &=& -1 + \frac{\alpha_s}{\pi} \left(
    \frac23\ln\frac{m_b}{2\mu_c} - \frac{20}{3}\,\zeta
    + \frac{7}{18} + \frac{8\pi^2}{9} \right) + \dots \,,
\end{eqnarray}
in agreement with the findings of \cite{Neubert:2000ch,Jeza}. The 
numerical effects of the NLO resummation of logarithms achieved in 
\cite{Neubert:2000ch} and in the present work are rather significant, 
despite the fact that there is limited phase space for operator 
evolution between the $b$-quark mass and the 
scale $\mu_c\approx 1$\,GeV. In the 
pole scheme (i.e., with $\zeta=0$) and with $q_0^2=(M_B-M_D)^2$, the 
running from $\mu=m_b$ down to a characteristic scale $\mu=\mu_c$ 
decreases the value of $\widehat\Gamma_3$ by a factor 
0.47, and it changes the value of $\widehat\Gamma_4$ by a factor 
$-0.49$. The resummation effects are less pronounced in the PS 
scheme with $\zeta=1$, where we find that the value of $\widehat\Gamma_3$
increases by a factor 1.10, whereas the value of 
$\widehat\Gamma_4$ decreases by a factor 0.66.

Finally, we quote the expression for the remainder 
$\widehat\Gamma_{\ge 5}$ in (\ref{hybrid}), which contains all terms 
suppressed by at least five powers of $\delta$, and for which 
we do not perform a RG improvement. As we will see below these 
contributions are numerically very small, so that a leading-order 
treatment is indeed justified. We obtain
\begin{equation}\label{remainder}
   x^{-12/25}\,\widehat\Gamma_{\ge 5}
   = \frac{\alpha_s(\bar\mu)}{\pi}\,G(\delta,\zeta)
   + (1-\delta)\,\frac{\lambda_1+15\lambda_2(\bar\mu)}{2\mu_c^2}
   + \dots \,,
\end{equation}
where the scale $\bar\mu$ is undetermined and will be chosen between
$\mu$ and $m_b$ in our analysis below. The function $G(\delta,\zeta)$ 
can be obtained from the results of \cite{Jeza} and is given by
\begin{eqnarray}
   G(\delta,\zeta) &=& \frac{1}{12\delta^5} \left[ L_2(1-2\delta)
    - \frac{\pi^2}{6} - \left( \frac{13}{12} - 8\delta + 28\delta^2
    - \frac{128}{3}\,\delta^3 + 20\delta^4 \right) \ln(1-2\delta)
    \right] \nonumber\\
   &&\mbox{}- \frac{1}{6\delta^4} \left[ \frac{1}{12}
    - \frac{89}{12}\,\delta + 21\delta^2 + \frac{25}{3}\,\delta^3
    + \left( 1 + \delta + \frac43\,\delta^2 - 14\delta^3 \right)
    \ln2\delta \right] \nonumber\\
   &&\mbox{}+ \frac{4(1-\delta)}{3\delta^2} \left[ L_2(2\delta)
    - L_2(1-2\delta) + \frac{\pi^2}{6} \right] + 4\zeta \nonumber\\
   &=& \frac{8}{15}\ln 2\delta + 4\zeta - \frac{544}{225} 
    - \frac{76}{135}\,\delta + O(\delta^2) \,.
\end{eqnarray}
For the optimal choice $q_0^2=(M_B-M_D)^2$, corresponding to the largest
value of $\delta$, and in the PS scheme with $\zeta\approx 1$ and 
$\mu_c<\mu<2\mu_c$, we find that the relative contributions of the three
terms involving $\widehat\Gamma_3$, $\widehat\Gamma_4$, and 
$\widehat\Gamma_{\ge 5}$ in (\ref{hybrid}) are approximately 
$1:-0.15:0.05$, indicating a rapid convergence of the hybrid expansion, 
and supporting our claim that resummation effects are unimportant for 
the quantity $\widehat\Gamma_{\ge 5}$, especially since more than 80\% 
of the contributions to this quantity come from power corrections.

\section{Numerical analysis}

We now turn to the numerical evaluation of our result for the decay rate
$\Gamma(\delta)$ 
and investigate the perturbative uncertainties in the calculation. The 
uncertainties related to higher-order power corrections 
$\sim(\Lambda/\mu_c)^3$ have been estimated previously 
\cite{Bauer:2000xf,Neubert:2000ch}.

The value of the PS mass at the subtraction scale $\mu_2=2$\,GeV has 
been determined using a sum-rule analysis of the $b\bar b$ production 
cross section near threshold \cite{Signer}. The result is
$\mbPS(\mu_2)=(4.59\pm 0.08)$\,GeV, which corresponds to
$\overline{m}_b(m_b)=(4.25\pm 0.08)$\,GeV in the $\overline{\rm MS}$
scheme. The most recent value for the Upsilon mass as reported in the
second paper in \cite{Andre} is $\mbUp=(4.68\pm 0.05)$\,GeV and 
corresponds to $\overline{m}_b(m_b)=(4.16\pm 0.06)$\,GeV. This paper
also contains a sum-rule analysis of the $b\bar b$ threshold region, 
yielding $\overline{m}_b(m_b)=(4.17\pm 0.05)$\,GeV. Whereas charm-quark
mass effects were included in \cite{Andre} they were neglected in 
\cite{Signer}. When comparing the results for the $B\to X_u\,l\,\nu$ 
decay rate obtained using different mass definitions, one should use
numerical values that correspond to the same value of the 
$\overline{\rm MS}$ mass. Combining the values 
reported above, and applying a small charm-quark mass correction in 
the case of \cite{Signer}, we choose 
$\overline{m}_b(m_b)=(4.20\pm 0.06)$\,GeV as a reference value. The
correspondingly shifted values for the low-scale subtracted quark masses 
are
\begin{equation}
   \mbPS(2\,\mbox{GeV}) = (4.52\pm 0.06)\,\mbox{GeV} \,, \qquad
   \mbUp = (4.72\pm 0.06)\,\mbox{GeV} \,.
\end{equation}
Below we will indicate how our results would change if different mass 
values were used. 

Given a value of $\mbPS(\mu_2)$, we use (\ref{eq:PSmass}) to compute the 
running PS mass at the scale $\mu_f=\zeta\mu_c$ by solving the NLO 
equations
\begin{eqnarray}
   \mbPS(\mu_f)
   &=& \mbPS(\mu_2) + \mu_f\,\frac{4\alpha_s(\mu_2)}{3\pi} \left[ 
    \left( \frac{\mu_2}{\mu_f} - 1 \right) \left( 1 + 
    \frac{203}{36} \frac{\alpha_s(\mu_2)}{\pi} \right)
    - \frac{25}{6} \frac{\alpha_s(\mu_2)}{\pi}
    \ln\frac{\mu_2}{\mu_f} \right] , \nonumber\\
   \mbPS(\mu_f) &=& \mu_c + \sqrt{\mu_c^2+q_0^2} \,.
\end{eqnarray}
For instance, with the optimal choice $q_0^2=(M_B-M_D)^2$ we obtain
$\mu_c\simeq 1.09$\,GeV, $\mbPS\simeq 4.67$\,GeV and 
$\delta\simeq 0.23$ for $\zeta=1$. The next question that arises is 
which values of the parameter $\zeta$ we should consider. A priori any 
value $\zeta=O(1)$ is reasonable, but in practice the choice 
is constrained by the following considerations. First, it is known that
the characteristic scale for the total $B\to X_u\,l\,\nu$ decay rate, 
while parametrically of order $m_b$, it is numerically much smaller than
the $b$-quark mass, of order 1--2\,GeV. 
A theoretical argument supporting this 
finding has been presented in \cite{BSUV97}. Because the cut on the 
dilepton invariant mass reduces the phase space for the hadronic 
invariant mass and energy, the characteristic scale for the decay rate 
$\Gamma(\delta)$ is lower than that for the total rate. Hence, the 
largest reasonable value of $\zeta$ is about 2, corresponding to a 
subtraction scale $\mu_f=2\mu_c\approx 2$\,GeV. Likewise, we should not 
consider renormalization scales larger than about 2\,GeV. This conclusion
is also borne out by closer inspection of the perturbative expansion 
coefficients in (\ref{master}). Whereas for $\zeta\approx 1$ the 
coefficients of the various terms proportional to $\alpha_s(\mu)$ in the 
expressions for $\widehat\Gamma_3$ and $\widehat\Gamma_4$ take moderate 
values (and are of much smaller magnitude than in the pole scheme), some 
of these coefficients become large for $\zeta\ge 2$. On the other hand, 
$\zeta$ can also not be taken too small. The formal limit $\zeta\to 0$ 
corresponds to the pole scheme and reintroduces the bad perturbative 
behavior of expansions based on using the pole mass. In addition, there 
are significant uncertainties in the calculation of the running PS mass 
when the subtraction scale is much below 1\,GeV, where NNLO 
corrections become important. Based on these observation we will employ 
the PS mass subtracted at the scale $\mu_f=\mu_c$ ($\zeta=1$) as our 
default choice, and consider the choices $\zeta=2$ and $\zeta=0.5$ as 
extreme variations.
 
\begin{figure}[t]
\begin{center}
\includegraphics[width=0.7\textwidth]{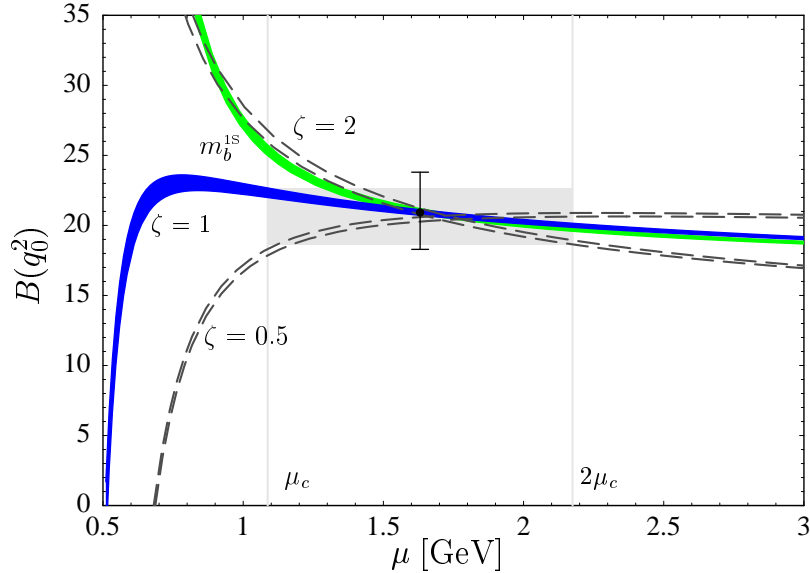}
\parbox{14cm}{\caption{\label{fig:mudep}
Scale dependence of the $B\to X_u\,l\,\nu$ branching ratio in units of
$|V_{ub}|^2$ for $q_0^2=(M_B-M_D)^2$, and for 
different mass definitions. The two bands refer to the PS mass 
subtracted at $\mu_f=\mu_c$ (dark), and to the Upsilon mass (light). The 
dashed lines refer to extreme choices of the subtraction scale for the 
PS mass. The gray rectangle shows our estimate of the perturbative 
uncertainty (see text for further explanation).}}
\end{center}
\end{figure}

To present our results for the branching ratio we define a function
$B(q_0^2)$ such that
\begin{equation}
   \mbox{Br}(B\to X_u\,l\,\nu)|_{q^2>q_0^2}
   \equiv B(q_0^2)\times |V_{ub}|^2
   \left( \frac{\tau_B}{1.6\,\mbox{ps}} \right) \,,
\end{equation}
where $\tau_B$ is the $B$-meson lifetime. In Figure~\ref{fig:mudep}, we 
show the renormalization-scale dependence of $B(q_0^2)$ for the optimal
choice $q_0^2=(M_B-M_D)^2$ and different mass definitions. We scan the 
parameters $\lambda_1$ and $\bar\mu$ entering the expression for 
$\widehat\Gamma_{\ge 5}$ in (\ref{remainder}) in the ranges 
$-0.45\,\mbox{GeV}\le\lambda_1\le-0.15$\,GeV and $\mu\le\bar\mu\le m_b$. 
The sensitivity to the values of these parameters is very small, as 
indicated by the widths of the bands. The vertical gray lines 
indicate the ``perturbative window'' between $\mu=\mu_c$ and $2\mu_c$ 
(calculated with $\zeta=1$). In this range there are no large 
logarithms, and the coupling $\alpha_s(\mu)$ is reasonably small. We 
observe good stability of the results in this region, and an excellent 
agreement between the PS scheme and the Upsilon scheme for scales above 
1.7\,GeV. For lower scales, the results in the Upsilon scheme become 
unstable, whereas the stability interval for the PS scheme extends 
down to $\mu<1$\,GeV. 
Remarkably, the curves corresponding to the different 
mass definitions all cross at a point close to the center of the 
window. We take the central value of the band corresponding to the PS 
mass with $\zeta=1$, evaluated at the center of the window, as our 
central value, and use the results obtained with $\zeta=1$, $\mu=\mu_c$ 
and $\zeta=2$, $\mu=2\mu_c$ to get an estimate of the perturbative 
uncertainty. 

The error bar at the default point reflects the sensitivity to the value
of the $b$-quark mass. The branching ratio scales as approximately the 
tenth power of 
$m_b$ \cite{Neubert:2000ch}, and because of this strong dependence this 
is the dominant theoretical uncertainty in our prediction. The dependence
on the $b$-quark mass can be parameterized as
\begin{equation}
   B(q_0^2)\propto 
   \left( \frac{\mbPS}{4.52\,\mbox{GeV}} \right)^{\Delta(q_0^2)} ,
\end{equation}
where $\mbPS\equiv\mbPS(2\,\mbox{GeV})$, and the exponent $\Delta(q_0^2)$
increases with $q_0^2$ (see Table~\ref{tab:results} below). 
Alternatively, one 
may substitute the ratio $\mbPS/4.52$\,GeV with $\mbUp/4.72$\,GeV or 
$\overline{m}_b(m_b)/4.20$\,GeV. The $\pm 60$\,MeV error assumed in the
figure is less conservative than the $\pm 80$\,MeV advocated in 
\cite{Martin}, but larger than the $\pm 50$\,MeV found in \cite{Andre}.

\begin{table}[t]
\centerline{\parbox{14cm}{\caption{\label{tab:results} 
Theoretical predictions for the branching ratio for different values of 
$q_0^2$. The optimal choice of the cutoff is $q_0^2=(M_B-M_D)^2\simeq 
11.6$\,GeV$^2$. Errors are added in quadrature and symmetrized in the 
final result.}}}
\vspace{0.2cm}
\begin{center}
{\tabcolsep=0.5cm
\begin{tabular}{|c|c|ccc|c|}
\hline\hline
$q_0^2$ [GeV$^2$] & $B(q_0^2)$ & $\delta m_b$ & pert.\
 & $(\Lambda/\mu_c)^3$ & $\Delta(q_0^2)$ \\
\hline
10.5 & $26.9\pm 4.5$ & ${}_{\,-3.0}^{\,+3.3}$ & ${}_{\,-3.0}^{\,+2.1}$
 & $\pm 2.0$ & $\phantom{1}8.8$ \\
$(M_B-M_D)^2$ & $20.9\pm 4.0$ & ${}_{\,-2.6}^{\,+2.9}$
 & ${}_{\,-2.3}^{\,+1.8}$ & $\pm 2.0$ & \phantom{1}9.9 \\
13.0 & $14.8\pm 3.4$ & ${}_{\,-2.2}^{\,+2.4}$ & ${}_{\,-1.5}^{\,+1.3}$
 & $\pm 2.0$ & 11.7 \\
15.0 & $\phantom{2}7.6\pm 2.7$ & ${}_{\,-1.6}^{\,+1.8}$
 & ${}_{\,-0.5}^{\,+0.7}$ & $\pm 2.0$ & 16.4 \\
\hline\hline
\end{tabular}}
\end{center}
\end{table}

Table~\ref{tab:results} shows our final theoretical predictions for 
$B(q_0^2)$ as a function of the cutoff on the dilepton invariant 
mass. We separately list the errors due to the uncertainty in the 
$b$-quark mass, the perturbative uncertainty, and higher-order power 
corrections. Each of the quantities $\widehat\Gamma_n$ in (\ref{hybrid}) 
receives corrections of order $(\Lambda/\mu_c)^3$, whose contributions 
are essentially unknown. Because of the prefactor 
$\delta^3=(\mu_c/m_b)^3$ in this equation, the corresponding contribution 
to the decay rate is to first order independent of $q_0^2$ and scales 
like $(\Lambda/m_b)^3$. Correspondingly we assign an error of
$\pm 16\Gamma_0\,(\Lambda/m_b)^3\approx 2\,|V_{ub}|^2$, where we have
used $\Lambda=500$\,MeV as a typical hadronic scale. Recently the 
possibility has been entertained that it might be possible to lower 
the $q^2$ cutoff below the charm threshold, because the contributions 
from low-lying charm states can be subtracted using experimental data 
\cite{Bauer:2000zb}. We have therefore included a result for 
$q_0^2=10.5$\,GeV$^2$ in the table.

The results in Table~\ref{tab:results} compare well with the ones 
obtained in \cite{Neubert:2000ch}, where a value corresponding to
$B(q_0^2)=20.0\pm 3.9$ was obtained for $q_0^2=(M_B-M_D)^2$. 
A new element of the present analysis is the study of different mass 
definition schemes, which allows us to obtain a more reliable estimate 
of the perturbative uncertainty. The authors of 
\cite{Bauer:2000xf} have calculated the decay rate in the Upsilon scheme 
using fixed-order perturbation theory at the scale $\mu=m_b$. With this
choice of scale and scheme we find $B(q_0^2)=17.8$, significantly less 
than the central value shown in the table.
These authors have also included a partial calculation of 
higher-order corrections (terms of order $\beta_0\alpha_s^2$) and used 
$\mbUp=M_{\Upsilon(1S)}/2$ for the Upsilon mass (with no uncertainty),
thereby neglecting the nonperturbative contribution estimated in 
\cite{Andre}. Their result reported in \cite{Bauer:2000xf} corresponds 
to $B(q_0^2)=(19.3\pm 0.8\,[\mbox{pert}]\pm 1.7\,[\mbox{power}])$. 
Since the $\beta_0\alpha_s^2$ terms are related to the choice of scale 
in the leading-order correction, it is not clear to us why the quoted 
perturbative error is smaller than the contribution of these terms. 
More importantly, however, the dominant theoretical uncertainty due to 
the strong sensitivity to the $b$-quark mass has been neglected in 
\cite{Bauer:2000xf}.

\section{Conclusions}

We have presented a calculation of the branching ratio for the inclusive
decay $B\to X_u\,l\,\nu$ with a cut $q^2>q_0^2\ge(M_B-M_D)^2$ on
the dilepton invariant mass. Since the typical parton
momenta $\sim\mu_c$ after the cut are of order the charm-quark
mass, we have performed a two-step expansion in the ratios
$\mu_c/m_b$ and $\Lambda/\mu_c$ and summed logarithms of the form 
$\alpha_s\ln(\mu_c/m_b)$ at next-to-leading order.
To improve the quality of the perturbative result, we have eliminated
the pole mass in favor of a low-scale subtracted $b$-quark mass. We
find that both the potential subtracted and the Upsilon mass lead
to consistent results for the rate. Considering the residual 
renormalization-scale dependence and the variations between different 
definitions of
the heavy-quark mass, we estimate the perturbative uncertainty to be
of order 10\%. This is comparable to the size of unknown 
higher-order power corrections, but smaller than the error arising from 
the uncertainty on the value of the $b$-quark mass.

If a value of the cutoff not far above the optimal choice
$q_0^2=(M_B-M_D)^2$ can be achieved experimentally, we conclude that
$|V_{ub}|$ can be determined with a precision of about 10\%. This
would be considerably less than the current theoretical uncertainty in
the value of this important Standard Model parameter.

\vspace{0.3cm}\noindent
{\it Acknowledgements:\/}
We are grateful to Martin Beneke for helpful discussions. This work was 
supported in part by the National Science Foundations of the U.S. and 
Switzerland.

\end{document}